%% file: paper420.tex
\documentclass[runningheads]{llncs}
\usepackage{graphicx}
\usepackage{amsmath}
\usepackage{amssymb}
\usepackage{booktabs}
\usepackage{multirow}
\DeclareMathOperator*{\argmin}{\arg\min}

\usepackage{caption}
\usepackage{subcaption}
\usepackage{bbding}

\usepackage[acronym]{glossaries}
\newacronym{dl}{DL}{Deep Learning}
\newacronym{vn}{GraDIRN}{Gradient Descent Network for Image Registration}
\newacronym{cnn}{CNN}{Convolutional Neural Networks}
\newacronym{ssd}{SSD}{sum of squared-difference}
\newacronym{cc}{CC}{cross-correlation}
\newacronym{ncc}{NCC}{normalized cross-correlation}
\newacronym{mi}{MI}{mutual information}

\usepackage{hyperref}
\hypersetup{breaklinks,colorlinks}

\usepackage[capitalize]{cleveref}
\crefname{section}{Sec.}{Secs.}
\Crefname{section}{Section}{Sections}
\Crefname{table}{Table}{Tables}
\crefname{table}{Tab.}{Tabs.}

\begin{document}

\title{Embedding Gradient-based Optimization in Image Registration Networks}
\titlerunning{Embedding Gradient-based Optimization in Image Registration Networks}

\author{Huaqi Qiu\inst{1}\textsuperscript{(\href{mailto:huaqi.qiu15@imperial.ac.uk}{\Envelope})}, Kerstin Hammernik\inst{1,2}, Chen Qin\inst{1,3}, Chen Chen\inst{1}, \\
Daniel Rueckert\inst{1,2}}
\institute{BioMedIA Group, Imperial College London, UK
\and  Klinikum rechts der Isar, Technical University of Munich, Germany
\and Institute for Digital Communications, University of Edinburgh, UK
\email{huaqi.qiu15@imperial.ac.uk}}
\authorrunning{Qiu H. et al.}

\maketitle              

\begin{abstract}
Deep learning (DL) image registration methods amortize the costly pair-wise iterative optimization by training deep neural networks to predict the optimal transformation in one fast forward-pass. 
In this work, we bridge the gap between traditional iterative energy optimization-based registration and network-based registration, and propose \acrfull{vn}. Our proposed approach trains a DL network that embeds unrolled multi-resolution gradient-based energy optimization in its forward pass, which explicitly enforces image dissimilarity minimization in its update steps.
Extensive evaluations were performed on registration tasks using 2D cardiac MR and 3D brain MR images. We demonstrate that our approach achieved state-of-the-art registration performance while using fewer learned parameters, with good data efficiency and domain robustness. 

\end{abstract}

\section{Introduction} \label{sec:intro}
Image registration aims to find the spatial transformation between corresponding anatomical or functional locations in different images. In medical imaging, image registration is one of the most fundamental tasks in image fusion, surgical intervention, motion analysis \cite{Bello2019-js} and disease progression analysis \cite{DBLP:journals/neuroimage/HuaLPLCTJWT08}. Many traditional registration methods find the transformation for each pair of images by minimizing an energy function using iterative optimization \cite{DBLP:journals/tmi/SotirasDP13}, often over multiple resolutions. The energy function usually consists of an image dissimilarity term, which enforces the alignment between the images, and a regularization term which imposes constraints on the transformation such as smoothness and invertibility. However, traditional iterative optimization is computationally demanding, especially for 3D images and high-dimensional transformations. 

More recently, data-driven methods perform image-to-transformation mapping in one step using deep neural networks,  which substantially speed up registration. Early \textit{supervised} methods~\cite{DBLP:conf/miccai/SokootiVBLIS17,DBLP:journals/neuroimage/YangKSN17} rely on ground truth transformation (parameters) as supervision. To avoid dependency on ground truth transformation, \textit{self-supervised} approaches~\cite{DBLP:journals/tmi/BalakrishnanZSG19,DBLP:conf/cvpr/MokC20,DBLP:journals/mia/VosBVSSI19} rely on dissimilarity metrics based on image intensities or semantic segmentation \cite{DBLP:journals/tmi/BalakrishnanZSG19,DBLP:conf/miccai/LeeOSSG19,DBLP:journals/mia/HuMGLGBWBMEONBV18}. 
Recent advances extend the one-step approach of regressing the transformation by refining the transformation over multiple resolutions and/or multiple steps~\cite{DBLP:conf/miccai/HeringGH19,Liu2021-la,DBLP:conf/nips/SandkuhlerABNJC19,DBLP:conf/cvpr/MokC20,DBLP:conf/iccv/ZhaoDCX19} analogous to traditional registration. However, these methods only enforce image dissimilarity minimization in the final training loss but not in the network forward pass steps.

Meanwhile, several methods in solving inverse problems such as image reconstruction and image de-noising~\cite{DBLP:journals/pami/ChenP17,DBLP:journals/tci/GiltonOW20,DBLP:journals/corr/HammernikKKRSPK17,DBLP:conf/dagm/KoblerKHP17} propose networks that unroll iterative optimization algorithms for a fixed number of iterations. These networks use learned regularization steps while explicitly enforcing data-consistency in their forward pass iterations and achieved state-of-the-art results.
Inspired by these approaches, we propose a novel DL registration method which embeds an unrolled multi-resolution gradient descent in the registration network, as shown in \cref{fig:vn}. Concretely, the transformation update steps in our approach make use of adaptive image dissimilarity gradient to explicitly enforce image dissimilarity minimization, while using generalized regularization update steps learned from training data.
Our main contributions in this work are as follows:
\begin{enumerate}
    \item We introduce \acrfull{vn} which connects traditional iterative energy optimization and modern network-based registration by embedding an unrolled multi-resolution gradient descent in the registration network;
    
    
    \item We extensively evaluate the efficacy of our proposed method which demonstrates state-of-the-art registration performance for both 2D intra-subject cardiac MR and 3D inter-subject brain MR registration tasks while being efficient in the number of learned parameters;
    
    \item We perform further evaluations under data scarce and domain shift settings, and demonstrate that the proposed method can be trained effectively with limited data while being robust to domain shift.
\end{enumerate}

\subsubsection{Related works} 
A few learning-based registration methods explored multi-step and multi-resolution iterative refinement of transformations~\cite{DBLP:conf/miccai/HeringGH19,Liu2021-la,DBLP:conf/cvpr/MokC20,DBLP:conf/miccai/XuCCCS21,DBLP:conf/iccv/ZhaoDCX19} to improve registration accuracy and handling of large deformations. These approaches essentially learn to regress the transformation via multiple network predictions via the training loss, but explicit image alignment is not enforced in these multiple steps.
Metric learning was proposed in \cite{DBLP:conf/cvpr/NiethammerKV19}, which uses a \acrshort{cnn} to learn a constrained multi-Gaussian local smoothing regularization applied on vector momentum stationary velocity field. In contrast, our proposed network learns the gradient update steps of the optimization instead of applying the learned regularization on the transformation. Hence, our method can be formulated with any transformation model and trained with standard deep learning training methods and does not require customized optimizers.
The variational registration network of~\cite{Jia2021-bv} uses linearized brightness consistency in learning an auxiliary-variable optimization, which limits the use of arbitrary metrics and transformation models. In contrast, our proposed method uses gradient descent optimization, which is easier to extend to other image dissimilarity metrics and transformation models, especially through our implementation using auto-differentiation (see \cref{method:lower}). In addition, unlike \cite{Jia2021-bv}, we use multi-resolution optimization with parameter-efficient networks, achieving competitive registration performance without ad-hoc learning of initialization.

\section{Method}
\begin{figure*}[t!]
     \centering
     \includegraphics[width=0.9\textwidth]{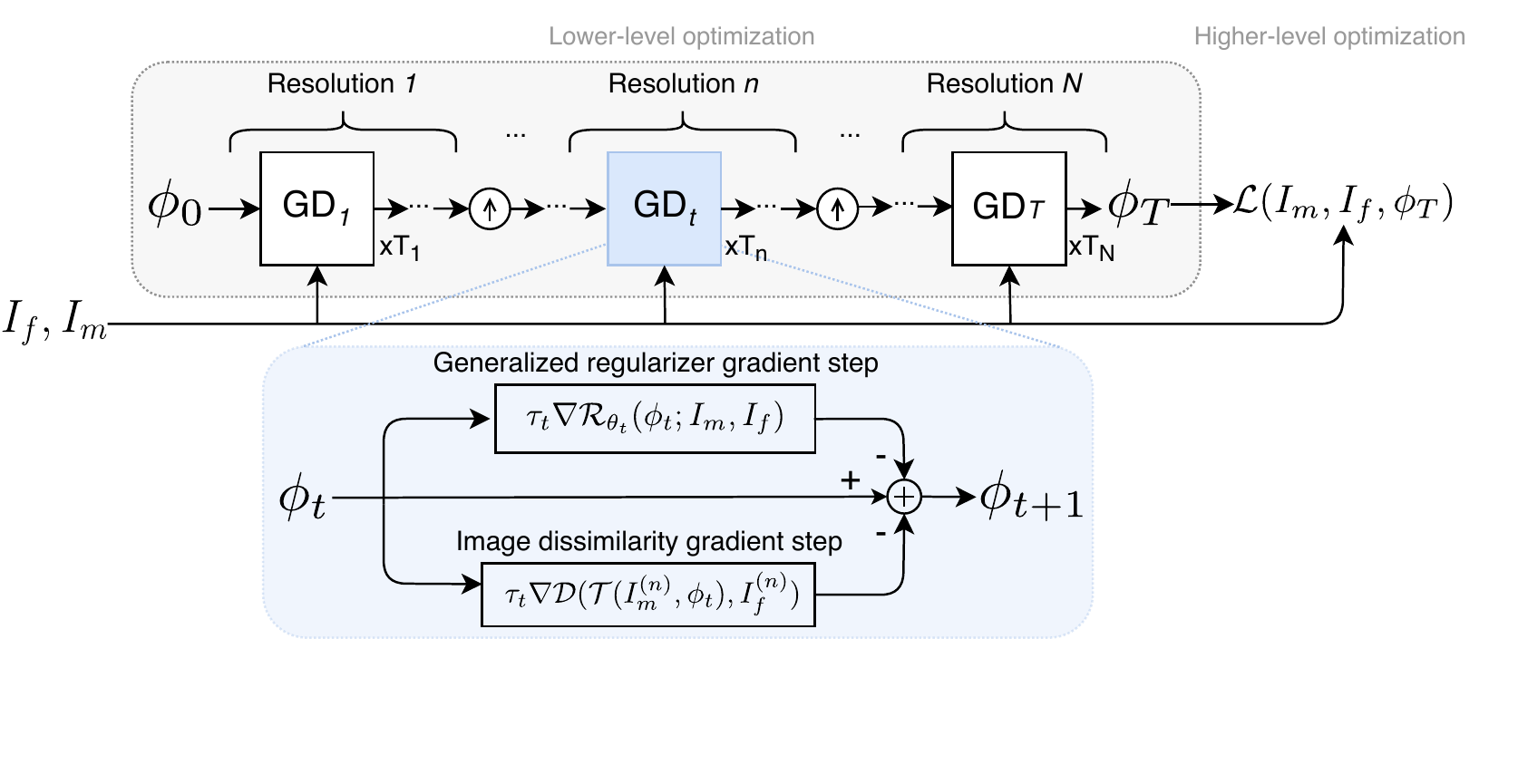}
     \caption{The proposed method under the bi-level optimization view. The lower-level optimization, shown in the gray box, is a multi-resolution optimization using explicit image dissimilarity gradient steps and \acrshort{cnn}-parameterized generalized regularizer steps. The blue block shows the inner structure of one gradient descent (GD) step. The higher-level optimization learns the parameters of the lower-level optimization by minimizing a loss function $\mathcal{L}$.}
     \label{fig:vn}
\end{figure*}

Our proposed image registration framework can be encapsulated in a bi-level optimization view, which is illustrated in~\cref{fig:vn}. The two levels of optimization are formulated as:

\begin{subequations}
    \begin{equation}\label{eq:higher}
        \min_{\theta} \mathbb{E}_{(I_m, I_f)\sim D} 
            [\mathcal{L}(I_m, I_f, \phi^*(\theta))]
    \end{equation}
    \begin{equation}\label{eq:lower}
        \text{s.t.} \ \  \phi^*(\theta) \in \argmin_{\phi} E(I_m, I_f, \phi).
    \end{equation}
\end{subequations}
%

%
The lower-level optimization in \cref{eq:lower} finds the optimal transformation $\phi^*$ to align a given pair of moving image $I_m$ and fixed image $I_f$ by minimizing the energy functional $E$.
We embed this optimization in the registration network by making the iterative steps that performs this optimization the forward pass of the network, which has trainable parameters $\theta$ (detailed in \cref{method:lower}).
Given the solution $\phi^*$ for each pair of images in the training dataset $D$, the higher-level optimization in \cref{eq:higher} finds the global optimal parameters for the network by minimizing the expectation of the loss function $\mathcal{L}( I_m, I_f,  \phi^*(\theta))$ (detailed in \cref{method:higher}).

\subsubsection{Lower-level optimization (forward pass)}\label{method:lower} 
To register a pair of images, the proposed network takes the image pair $(I_m, I_f)$ as input and predicts the transformation $\phi^*$, which is similar to existing registration networks (e.g. \cite{DBLP:journals/tmi/BalakrishnanZSG19}). However, the structure of our network is designed to solve the lower-level optimization problem which is the energy minimization in \cref{eq:lower}. The energy functional typically is of the form:
\begin{equation}\label{eq:energy}
    E(I_m, I_f, \phi) = \mathcal{D}(\mathcal{T}(I_m, \phi), I_f) + \alpha  \mathcal{R}(\phi).
\end{equation}
which contains an image dissimilarity $\mathcal{D}$ measured between the transformed moving image $\mathcal{T}(I_m, \phi)$ and the fixed image $I_f$, and a regularization term $\mathcal{R}$.
The proposed network unrolls the energy functional minimization using a gradient descent-based optimization with fixed number of update steps. An update step to $\phi_t$ at step $t$ given by:
\begin{equation} \label{eq:gd}
  \phi_{t+1} = \phi_t - \tau_t (\nabla\mathcal{D}(\mathcal{T}(I_{m}, \phi_t), I_f), +\nabla\mathcal{R}_{\theta_t}(\phi_t;I_m,I_f)),
\end{equation}

where $\tau_t$ is the adaptive step sizes, $\nabla \mathcal{D}$ is the image dissimilarity gradient and $\nabla\mathcal{R}_{\theta_t}$ is the trainable generalized regularization gradient functional, w.r.t. to the transformation $\phi$. This yields the network structure illustrated in \cref{fig:vn} with details explained in the following sections. 
\paragraph{Image Dissimilarity Gradient} 
Instead of only enforcing image alignment in the loss function as most existing DL network-based methods, we explicitly enforce image dissimilarity minimization in our network forward pass. The image dissimilarity gradient w.r.t. the transformation, $\nabla\mathcal{D}$, is directly applied in each update step. The forward computation of this gradient needs to be differentiable to allow back-propagation. To implement this, we leverage the power of modern auto-differentiation engines \cite{DBLP:conf/nips/PaszkeGMLBCKLGA19} to create the forward computation graph of $\nabla\mathcal{D}$ automatically. This enables us to use arbitrary dissimilarity metrics such as \acrfull{ssd} or \acrfull{ncc}, as well as other transformation models.
\paragraph{Learned Regularizer Gradient} 
For the regularizer gradient steps, we take inspiration from previous works~\cite{DBLP:journals/pami/ChenP17,DBLP:journals/corr/HammernikKKRSPK17} and use a \acrshort{cnn} to incorporate rich image information and learn more complex local regularization from training data. At each step $t$, the \acrshort{cnn} takes the current solution $\phi_t$ and the image pair $(I_m, I_f)$ as input, and outputs the gradient step corresponding to $\nabla\mathcal{R}_{\theta_t}$ in \cref{eq:gd}. We hypothesize that the use of explicit image dissimilarity gradient enables us to use more parameter-efficient networks in learning the \textit{regularizer} steps and optimal registration. So we use a simple 5-layer \acrshort{cnn} with 32 channels per layer, kernel size of 3 and LeakyReLU activation function. The formulation of our method is not limited to any specific network architecture. 
%
\paragraph{Multi-resolution optimization}
The transformation is optimized over multiple resolutions to efficiently capture both local deformation and global deformation. We initialize $\phi_0$ with identity transformation and update the transformation in a coarse-to-fine fashion. For resolution $n$, the transformation is updated $T_n$ times then upsampled to a higher resolution. We simply use a dense displacement field as the transformation model thus use linear interpolation for upsampling to achieve continuous optimization between resolutions. The final solution is $\phi^*=\phi_T$ where $T = \sum_{n=1,2,...N} T_n$. We share the parameters of the CNNs within each resolution which, together with the adaptive step sizes, are the learnable parameters $\theta=\{\theta_t,\tau_t\}_{t=1,2,...T}$ that need to be determined by the high-level optimization.

\subsubsection{Higher-level optimization (training)}\label{method:higher}
The learnable parameters $\theta$ in the lower-level optimization are learned in the higher-level optimization. This is achieved by minimizing the expected loss $\mathcal{L}$ over a training dataset $X$:
\begin{equation}\label{eq:loss}
 \mathcal{L}( I_m, I_f,  \phi^*(\theta)) = 
     \mathcal{L}_{\mathcal{D}}(\mathcal{T}(I_{m}, \phi^*(\theta)), I_f) +  \lambda\mathcal{L}_{\mathcal{R}}(\phi^*(\theta)).
\end{equation}
given the solution $\phi^*(\theta)$ from the lower-level optimization. $\mathcal{L}_\mathcal{D}$ is a \textit{self-supervised} loss using image dissimilarity, which is in the same mathematical form as the image dissimilarity term $\mathcal{D}$ in the energy function of the lower-level optimization for each training sample. The regularization loss we used is a standard diffusion penalty on the spatial variations of the transformation, namely $\mathcal{L}_{\mathcal{R}} = \frac{1}{|\Omega|} \sum_{\mathbf{x}\in \Omega}
\left\Vert\nabla\phi(\mathbf{x})\right\Vert_2^2$, with $\mathbf{x}$ denoting a point in the spatial domain $\Omega$. 
We use stochastic gradient-based optimization for the higher-level optimization using gradient of the loss function computed w.r.t. the learned parameters $\theta$ (not w.r.t. the transformation in \cref{eq:gd}). Since the lower-level optimization is unrolled into the forward pass of the network, the higher-level optimization is simply a standard deep learning training without the need for customized optimizers.

\section{Experiments}
\subsubsection{Tasks and Datasets}
We extensively evaluate registration performance on: 1) 2D intra-subject cardiac MRI registration between the end-diastolic (ED) frame and end-systolic (ES) frame; 2) 3D inter-subject brain MRI registration.
We use 220 cardiac cine-MR sequences from the publicly accessible UK Biobank (UKBB) study\footnote{\url{http://imaging.ukbiobank.ac.uk}}, with 100, 20 and 100 sequences used for training, validation and testing. We acquired semantic segmentation of the left ventricle cavity (LV), myocardium (Myo) and right ventricle cavity (RV) using a CNN-based segmentation model~\cite{Bai2018-yc} for evaluation purposes.
In addition, cardiac MRI of 150 subjects from the multi-vendor multi-center M\&Ms dataset~\cite{DBLP:journals/tmi/CampelloGIMSFMZ21} were used to evaluate domain generalization, with 75 subjects from each MR scanner vendor.
For the 3D brain MRI registration, we used T1-weighted MR images from the CamCAN open data inventory~\cite{Taylor2017-rd} which consist of structural brain MR scans of subjects aged between 18 and 90. The large structural variations in the dataset is challenging for inter-subject registration. We randomly select 200/10/100 scans for training/validation/testing. All scans are spatially aligned to a common MNI space via affine registration, skull-stripped using \href{https://www.nitrc.org/projects/robex}{ROBEX}, bias-field corrected using the N4 algorithm in \href{https://simpleitk.org}{SimpleITK}. Segmentation of 138 cortical and sub-cortical structures was acquired using a multi-atlas segmentation tool \href{https://github.com/ledigchr/MALPEM}{MALPEM}~\cite{Ledig2015-zv} for evaluation. 

\begin{table}[t!]
\centering
\caption{Evaluation results on the UKBB and CamCAN dataset reported in mean(standard deviation). The number of learnable parameters in the models are indicated by ``\#params". The best results in each column are in bold, * marks results not significantly different with the best results ($p\text{-value}>0.001$, one-sided Wilcoxon signed-rank test). Runtimes (CPU/GPU) are reported in seconds except 2D GPU in ms.}
\resizebox{\textwidth}{!}{%

\begin{tabular}{l|lllll|ll}
\hline
Dataset                                                                & Method        & Dice$\uparrow$        & HD$\downarrow$        & $|J|<0\% \downarrow$  & $\text{std}(\log(|J|)) \downarrow$ & Runtime     & \#params \\ \hline
\multirow{5}{*}{\begin{tabular}[c]{@{}l@{}}UKBB\\ (2D)\end{tabular}}   & Initial       & 0.500(0.058)          & 16.091(2.625)         & n/a                   & n/a                                & n/a          & n/a      \\
                                                                       & FFD           & 0.788(0.061)          & 10.970(3.006)         & 0.327(0.165)          & 1.093(0.317)                       & 1.534/-      & n/a      \\
                                                                       & VM            & 0.805(0.053)          & 10.243(2.725)         & 0.239(0.095)          & 0.976(0.208)                       & 0.011/2.796  & 106K     \\
                                                                       & RC-VM         & 0.820(0.052)*         & 9.900(2.636)          & 0.212(0.079)          & 0.923(0.181)                       & 0.101/59.502 & 954K     \\
                                                                       & \acrshort{vn} & \textbf{0.821(0.052)} & \textbf{9.729(2.531)} & \textbf{0.188(0.076)} & \textbf{0.866(0.182)}              & 0.075/59.502 & 88.5K    \\ \hline
\multirow{5}{*}{\begin{tabular}[c]{@{}l@{}}CamCAN\\ (3D)\end{tabular}} & Initial       & 0.621(0.043)          & 6.354(0.959)          & n/a                   & n/a                                & n/a          & n/a      \\
                                                                       & FFD           & 0.797(0.022)*         & \textbf{5.113(0.934)} & 0.084(0.060)          & 0.639(0.172)                       & 16.435/-     & n/a      \\
                                                                       & VM            & 0.780(0.026)          & 5.391(0.961)          & 0.060(0.050)          & 0.533(0.165)                       & 0.858/0.003  & 320K     \\
                                                                       & RC-VM         & 0.794(0.022)          & 5.179(0.903)*         & \textbf{0.047(0.041)} & \textbf{0.478(0.147)}              & 9.673/0.416  & 2883K     \\
                                                                       & \acrshort{vn} & \textbf{0.799(0.022)} & 5.147(0.857)*         & 0.056(0.050)          & 0.514(0.164)                       & 4.589/0.220  & 269K     \\ \hline
\end{tabular}%
}
\label{tab:main}
\end{table}

\subsubsection{Evaluation Metrics}\label{sec:metrics}
To comprehensively evaluate registration performances, we measure both the accuracy and the regularity of the transformation. In-lieu of ground truth transformation, we follow \cite{DBLP:journals/tmi/BalakrishnanZSG19} and measure registration accuracy based on the agreement of anatomical structures between the registered images using Dice score and Hausdorff Distance (HD). To evaluate the regularity of the transformations, we calculate the determinant of the Jacobian of the transformation $|J|=|\nabla \phi|$, and measure local folding indicated by  negative $|J|$ (reported in percentage denoted by $|J|<0\%$) and spatial smoothness by $\text{std}(\log |J|)$.

\subsubsection{Baselines and Implementation}
We first compare the proposed method to a traditional iteratively multi-resolution registration method which uses B-spline free-form deformation (FFD) and an energy functional consisting of an image dissimilarity metric and Bending Energy regularization \cite{DBLP:journals/tmi/RueckertSHHLH99}. We use the implementation in \href{https://mirtk.github.io}{Medical Image Registration ToolKit (MIRTK)} with three resolution levels and the same image dissimilarity metric as our method. We also compare our proposed method to two deep learning based registration methods. One is the widely used VoxelMorph framework~\cite{DBLP:journals/tmi/BalakrishnanZSG19} (VM). The other is a recursive-cascaded method introduced in \cite{DBLP:conf/iccv/ZhaoDCX19} (RC-VM) as a representative baseline for network-only (no explicit image matching) iterative DL registration, using VM as the base network for the cascade. We use 9 iterations ($T_N{=}9$) in our method (\acrshort{vn}) with 3 resolutions and 3 iterations per resolution ($T_n{=}3$). The number of cascades on RC-VM is set to be the same. The DL baselines are trained using the same loss function as the ones used in the higher-level optimization of the proposed method. All hyper-parameters are carefully tuned using the validation data set. 
Adam \cite{DBLP:journals/corr/KingmaB14} optimizer is used for training, with $\beta_1=0.9$ , $\beta_2=0.999$ and learning rate of $10^{-4}$. Our code is available at: \url{https://github.com/gradirn/gradirn}.



\subsubsection{Results and Discussion}
\begin{figure*}[t!]
     \centering
     \begin{subfigure}[h]{0.495\textwidth}
         \centering
         \includegraphics[width=\textwidth]{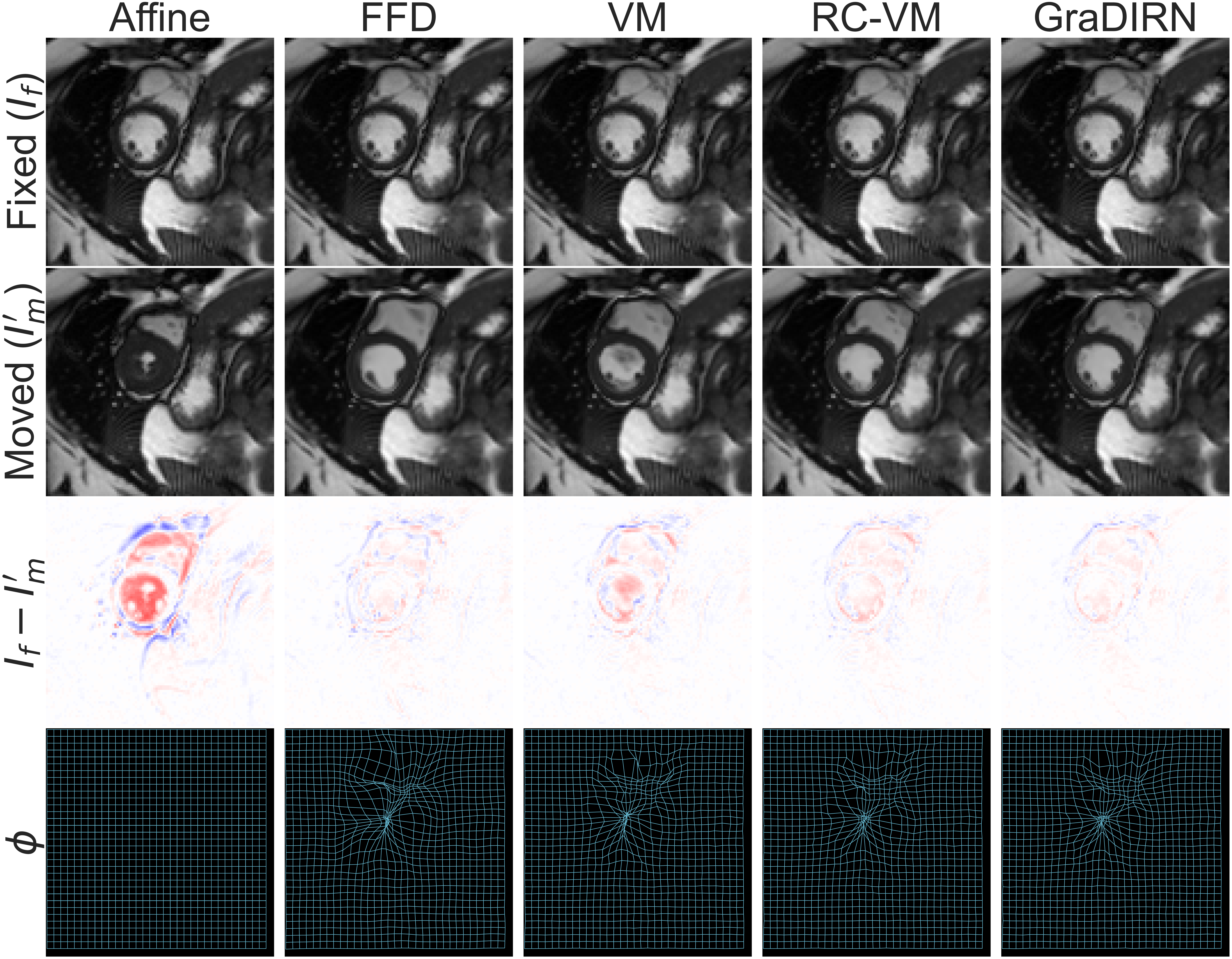}
         \label{image:cardiac}
     \end{subfigure}
     \begin{subfigure}[h]{0.495\textwidth}
         \centering
         \includegraphics[width=\textwidth]{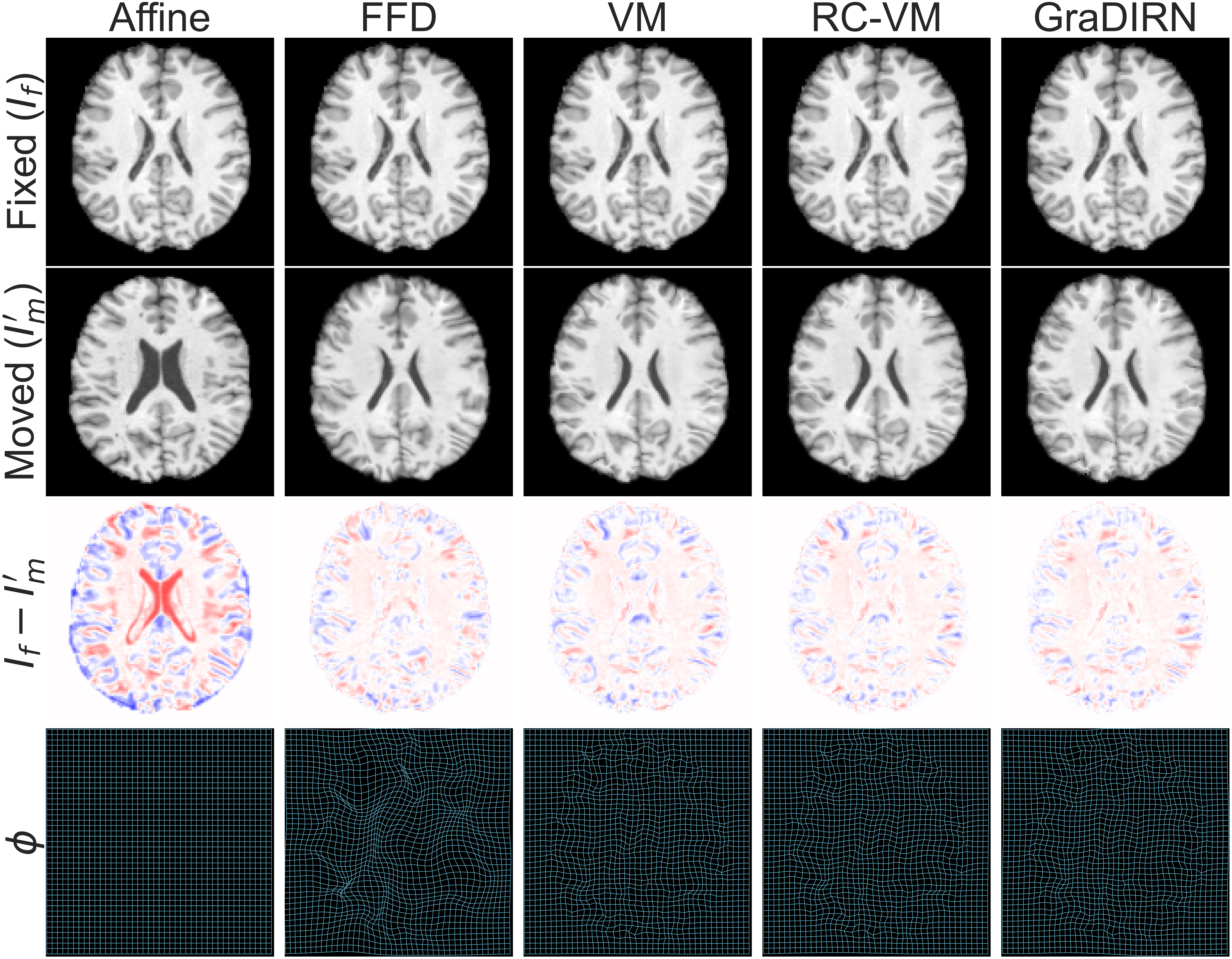}
         \label{img:brain}
     \end{subfigure}
     \caption{Examples of cardiac MR and brain MR registration results. A mid-ventricle slice is shown for the cardiac registration results and an axial view is shown for the brain registration results. The first row shows the fixed image $I_f$ (same for all methods), the second row shows the moving image transformed by the estimated transformation, the third row shows the relative intensity difference after registration, with the transformation shown in the final row.}
    \label{img:results}
\end{figure*}

\Cref{tab:main} shows the results of quantitative evaluation of all the competing methods using \acrshort{ncc} as the dissimilarity metric. On the 2D cardiac MR intra-subject registration task, the proposed method (\acrshort{vn}) significantly outperforms FFD and VM, comparable with RC-VM on Dice score but better on HD. On the 3D brain MR registration task, the proposed method results in higher Dice than the learning-based baselines and comparable performance as FFD, which is computationally much slower. The learning-based iterative approach RC-VM significantly improves VM on both registration tasks but uses networks with $~10\times$ larger capacity. In contrast, our proposed method is able to match the performance of RC-VM with even less learned parameters than VM.
Boxplots showing the Dice score for different anatomical structures can be found in the supplementary material. Visual examples of the registration results are shown in \cref{img:results}. 
We isolated and studied the effectiveness of the proposed use of the explicit dissimilarity gradient through ablation studies shown in \cref{tab:ablation}. It can be seen that the recursive cascade mechanism and \acrshort{vn} only with CNN updates perform inferior to \acrshort{vn} with the dissimilarity gradient updates.

\begin{figure}[t]
     \centering
     \begin{subfigure}[h]{0.23\textwidth}
         \centering
         \includegraphics[width=\textwidth]{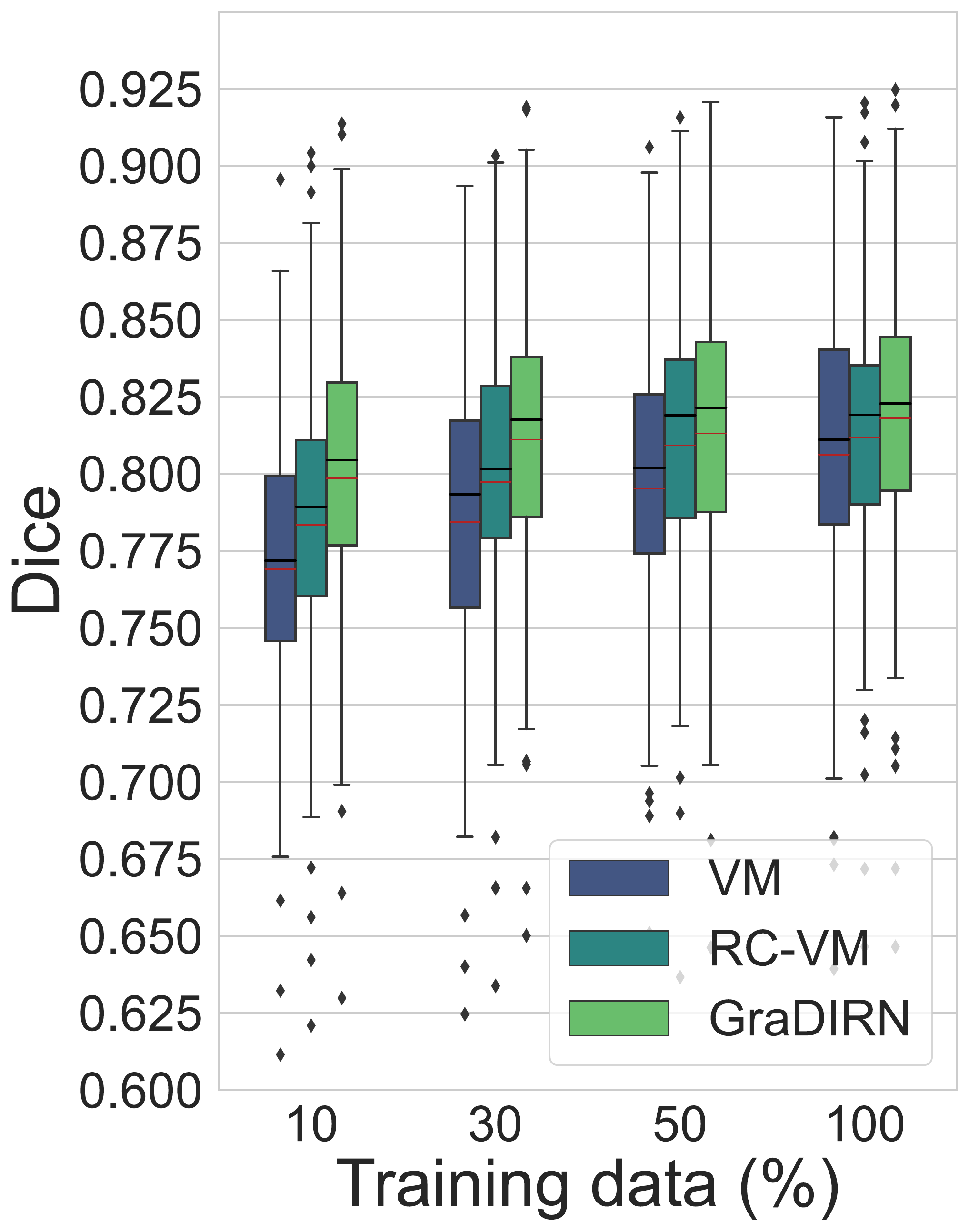}
         \caption{}
         \label{data:box}
     \end{subfigure}
     \begin{subfigure}[h]{0.23\textwidth}
         \centering
         \includegraphics[width=\textwidth]{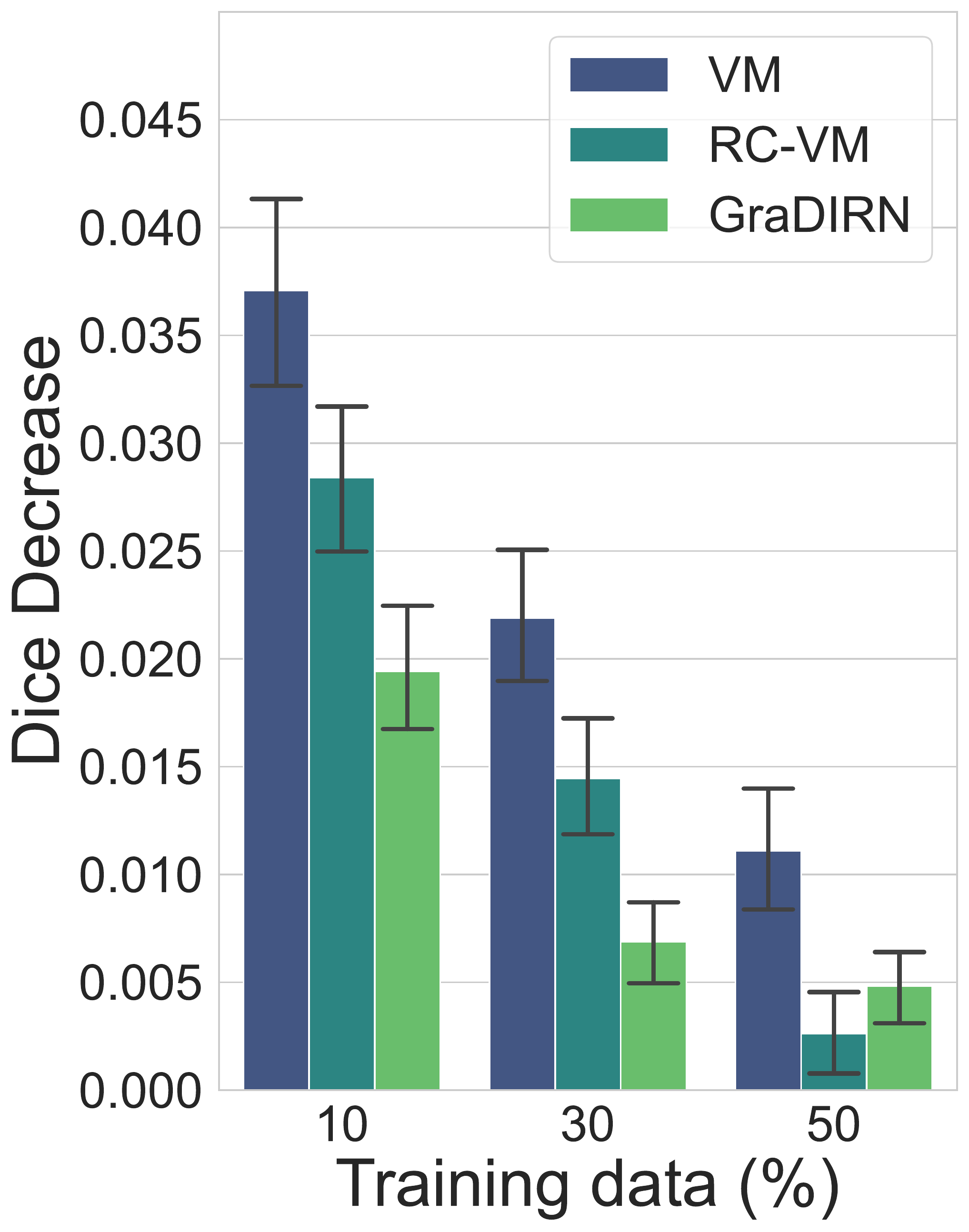}
         \caption{}
         \label{data:bar}
     \end{subfigure}
     \begin{subfigure}[h]{0.23\textwidth}
         \centering
         \includegraphics[width=\textwidth]{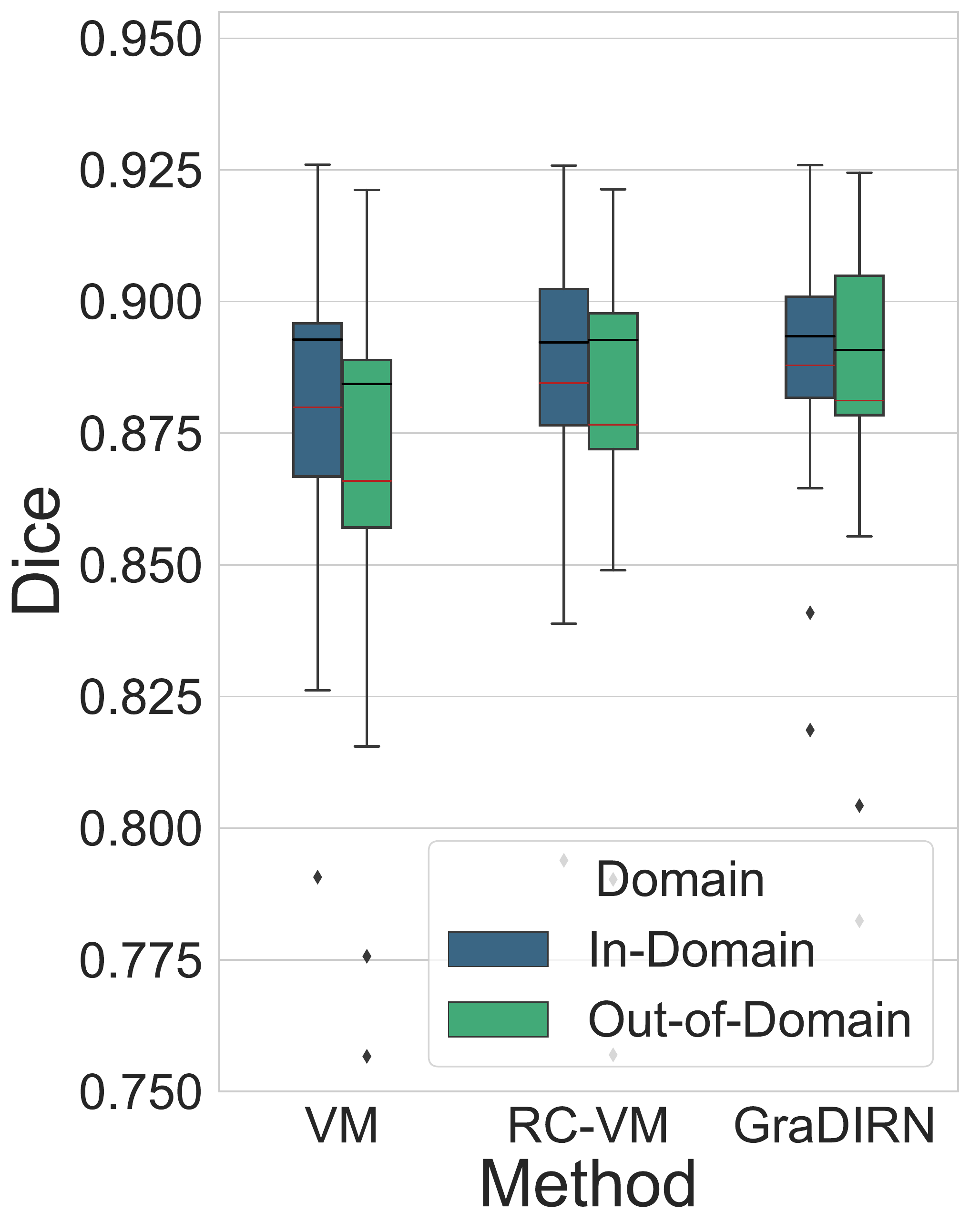}
         \caption{}
         \label{domain:box}
     \end{subfigure}
     \begin{subfigure}[h]{0.23\textwidth}
         \centering
         \includegraphics[width=\textwidth]{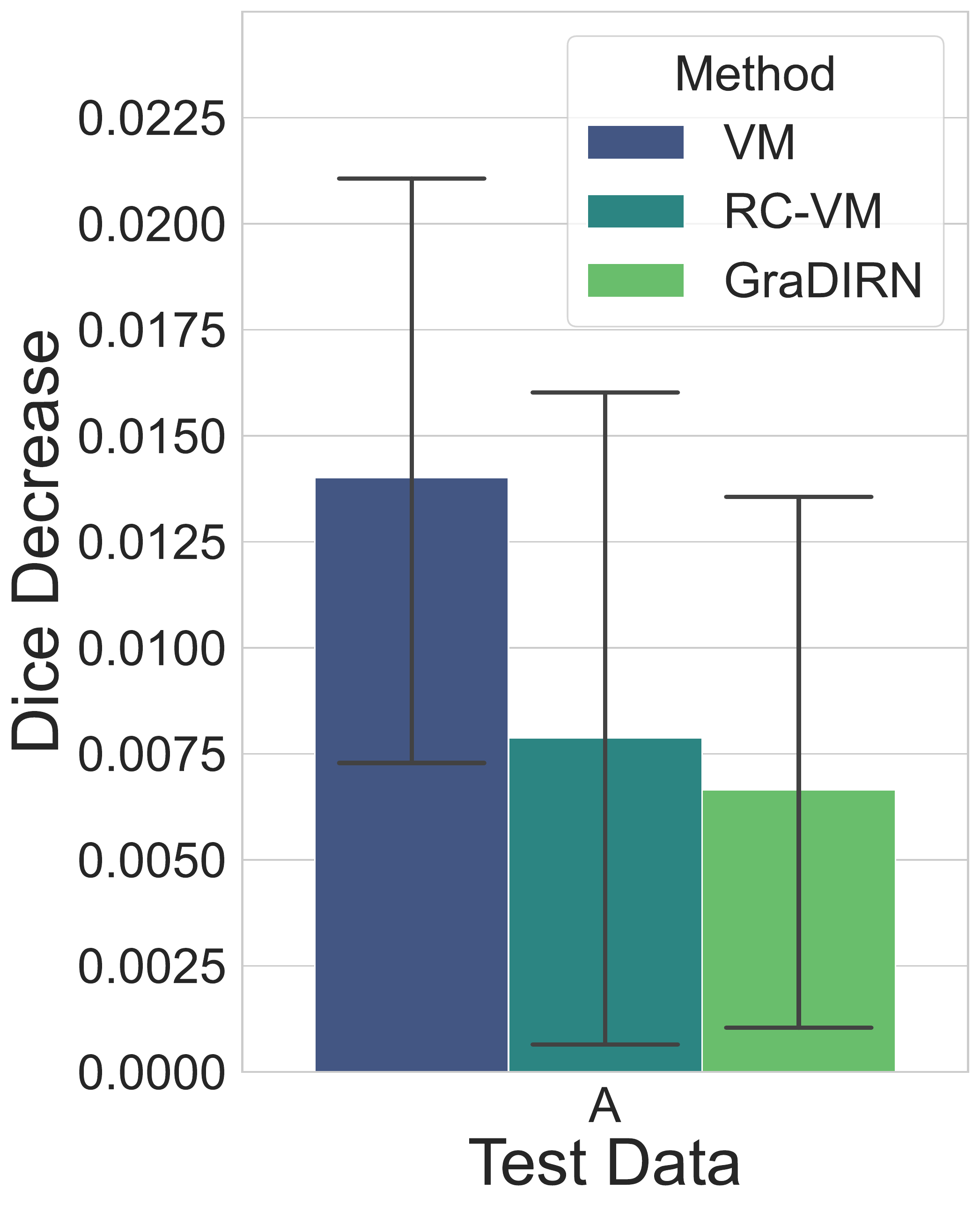}
         \caption{}
         \label{domain:bar}
     \end{subfigure}
     \caption{\textit{Data Efficiency: }\cref{data:box} and \ref{data:bar} show the performance and performance decrease of competing methods trained with limited data; \textit{Domain Robustness:} \cref{domain:box} and \cref{domain:bar} show the performance and performance difference of models tested on M\&Ms vendor A data but trained on vendor B (out-of-domain), compared to models trained on vendor A (in-domain).} 
\end{figure}

\paragraph{Data Efficiency}
We evaluate how the performance of competing method are affected when less training data is available. We trained all models with $10\%, 30\%$ and $50\%$ of the original UK biobank training dataset, and tested on the original testing dataset. The results, shown in \cref{data:box} and \cref{data:bar}, demonstrate that the proposed method can be more effectively trained with less data. With as little as $10\%$ of the original training data, our method can still achieve superior accuracy over the non-learning FFD.

\paragraph{Domain Robustness}
Learning-based registration methods are affected by domain shift between training and testing data but is rarely studied \cite{DBLP:conf/miccai/FerranteOGM18}. Therefore, we evaluate the \textit{domain robustness} of the learning-based methods using cardiac MR data from different vendors in the M\&Ms dataset \cite{DBLP:journals/tmi/CampelloGIMSFMZ21}. 
\cref{domain:box} and \cref{domain:bar} show the results of each model trained using the data from vendor B (Philips, in-domain) and tested on data from vendor A (Siemens, out-of-domain), which is a challenging cross-domain setting as reported in \cite{DBLP:journals/tmi/CampelloGIMSFMZ21}, compared to the same model trained using data from vendor A as reference. The proposed method shows better domain robustness, especially compared to VM, indicated by less Dice score drop and variation caused by domain shift.

\begin{table}[!tb]
\centering
\caption{Ablation study results (using \acrshort{ssd} for image dissimilarity). $\dagger$ indicates no image dissimilarity gradient. RC-CNN$^5$ is a variant of RC-VM using the same 5-layer CNN in our \acrshort{vn} as base network, where * indicate sharing every 3 cascades to match the number of parameters as \acrshort{vn}.} \label{tab:ablation}
\resizebox{0.7\textwidth}{!}{%
\begin{tabular}{l|lllll}
\hline
Settings                  & Dice         & HD            & $|J|<0\%$    & $std(\log(|J|))$ & \#params \\ \hline
RC-CNN$^5$          & 0.778(0.054) & 10.967(2.721) & 0.127(0.064) & 0.695(0.191      & 265K     \\
RC-CNN$^{5*}$ & 0.774(0.054) & 11.075(2.674) & 0.129(0.059) & 0.704(0.180)     & 88.5K    \\
\acrshort{vn}$\dagger$    & 0.799(0.052) & 10.481(2.738) & 0.082(0.050) & 0.547(0.182)     & 88.5K    \\
\acrshort{vn}             & 0.818(0.053) & 9.943(2.766)  & 0.073(0.041) & 0.510(0.163)     & 88.5K    \\ \hline
\end{tabular}%
}
\end{table}

\section{Conclusion}
In this paper, we present a novel learning-based registration network for deformable image registration which learns an unrolled multi-resolution gradient-based optimization with explicit image dissimilarity minimization embedded in the network forward pass.
Extensive evaluations show that our approach obtains state-of-the-art registration performance while retaining parameter efficiency, data efficiency and domain robustness. 
By using gradient-based optimization and auto-differentiation, our framework can easily incorporate arbitrary differentiable image dissimilarity metrics and transformation models with existing forward-pass implementation. 
\\

\noindent\textbf{Acknowledgments} 
This work was supported by the EPSRC Programme Grant EP/P001009/1. 
\clearpage

%
%

\bibliographystyle{splncs04}
\bibliography{paper420}


\appendix
\input{supplmentary}

\end{document}

%% file: supplmentary.tex
\begin{figure*}[h!]
     \centering
     \begin{subfigure}[h]{0.45\textwidth}
         \centering
         \includegraphics[width=\textwidth]{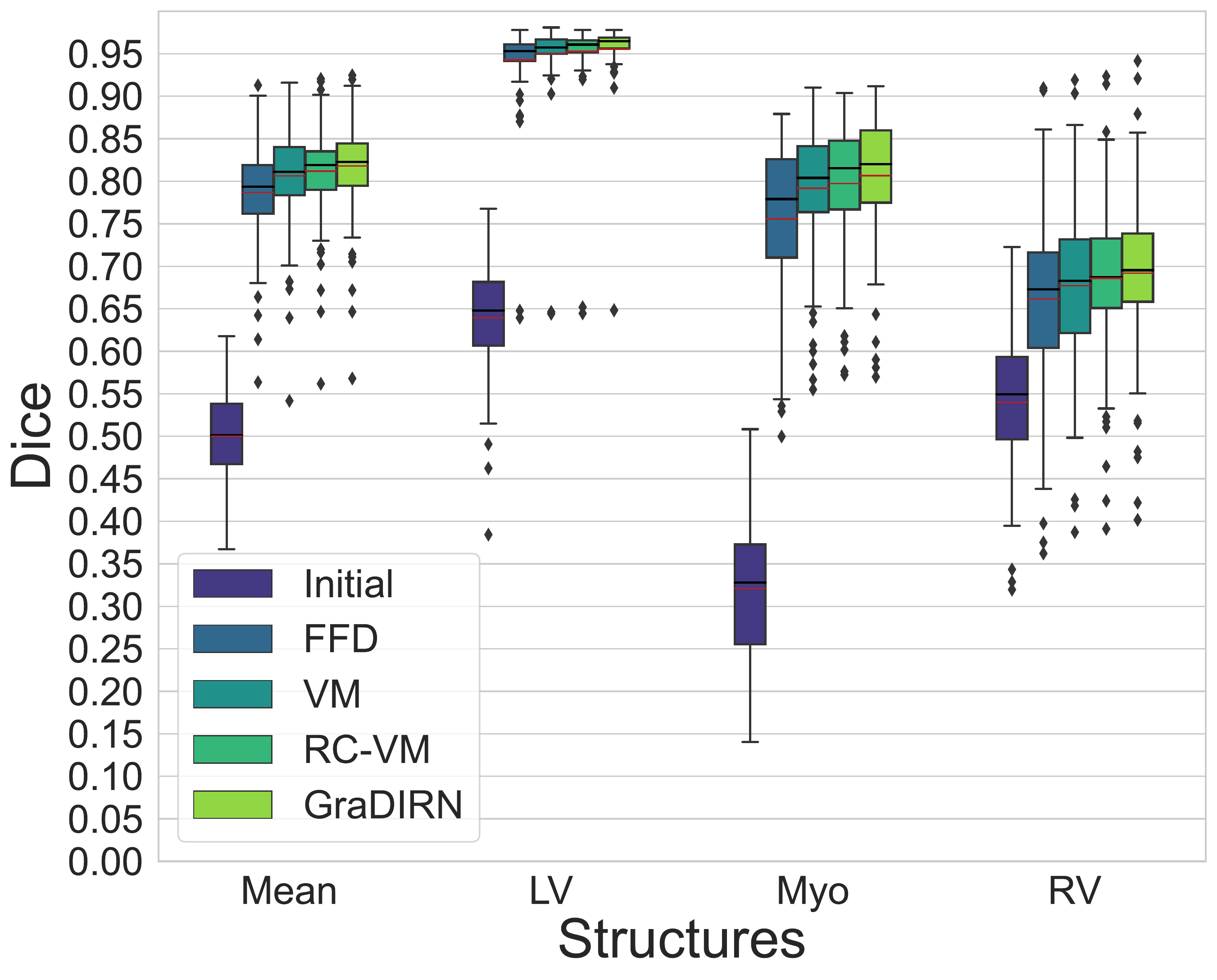}
         \caption{SSD}
     \end{subfigure}
     \hfill
     \begin{subfigure}[h]{0.45\textwidth}
         \centering
         \includegraphics[width=\textwidth]{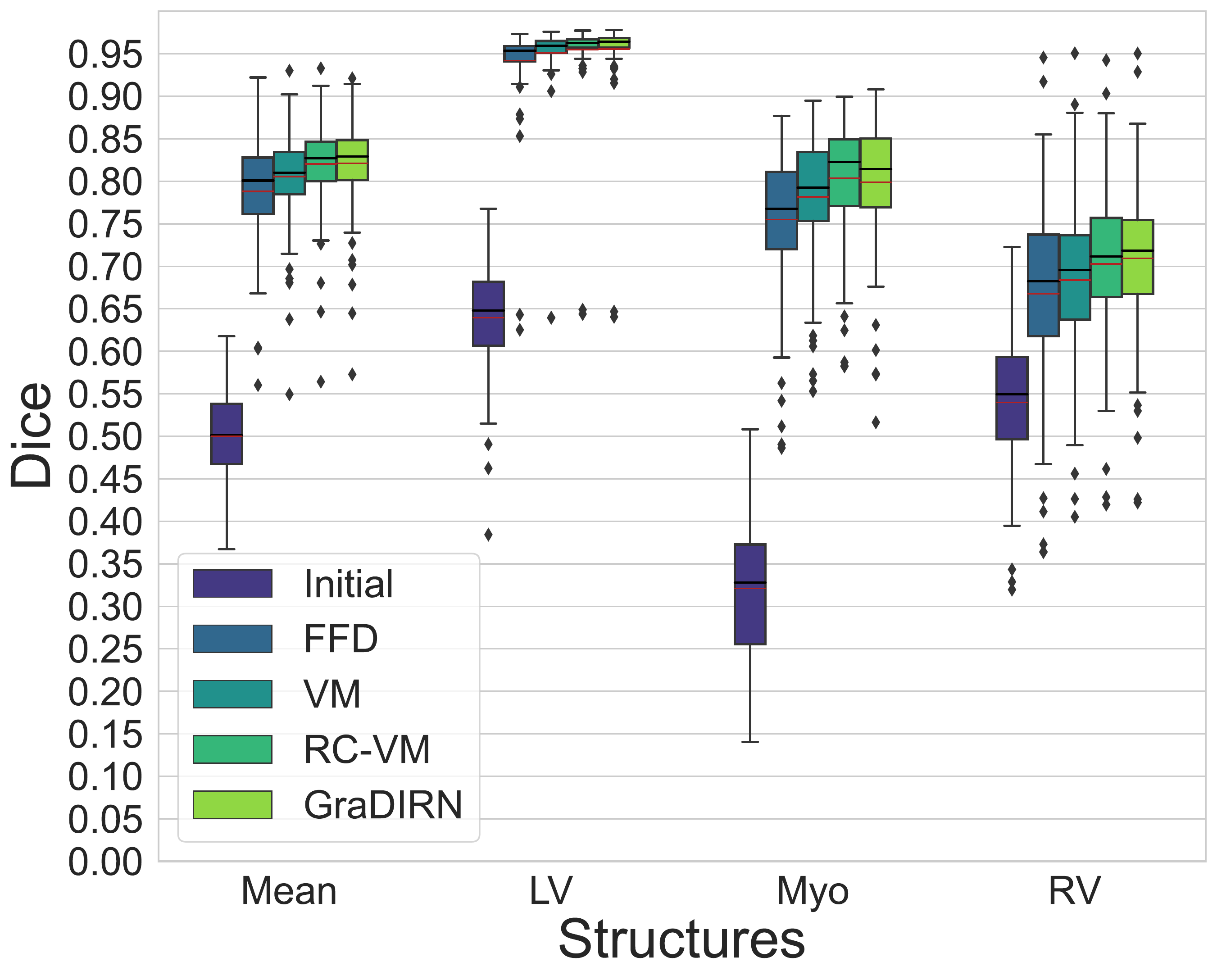}
         \caption{NCC}
     \end{subfigure}
     \caption{Boxplots of Dice scores measuring the overlap of different anatomical structures after registered by different methods in the cardiac registration task (UK biobank dataset). The black lines mark the medians and the red lines mark the means. Results using both sum-of-squared difference (SSD) and normalized cross-correlation (NCC) are shown.}
     \label{fig:cardiac_boxplots}
\end{figure*}
%
\begin{figure*}[h!]
     \centering
     \begin{subfigure}[h]{0.48\textwidth}
         \centering
         \includegraphics[width=\textwidth]{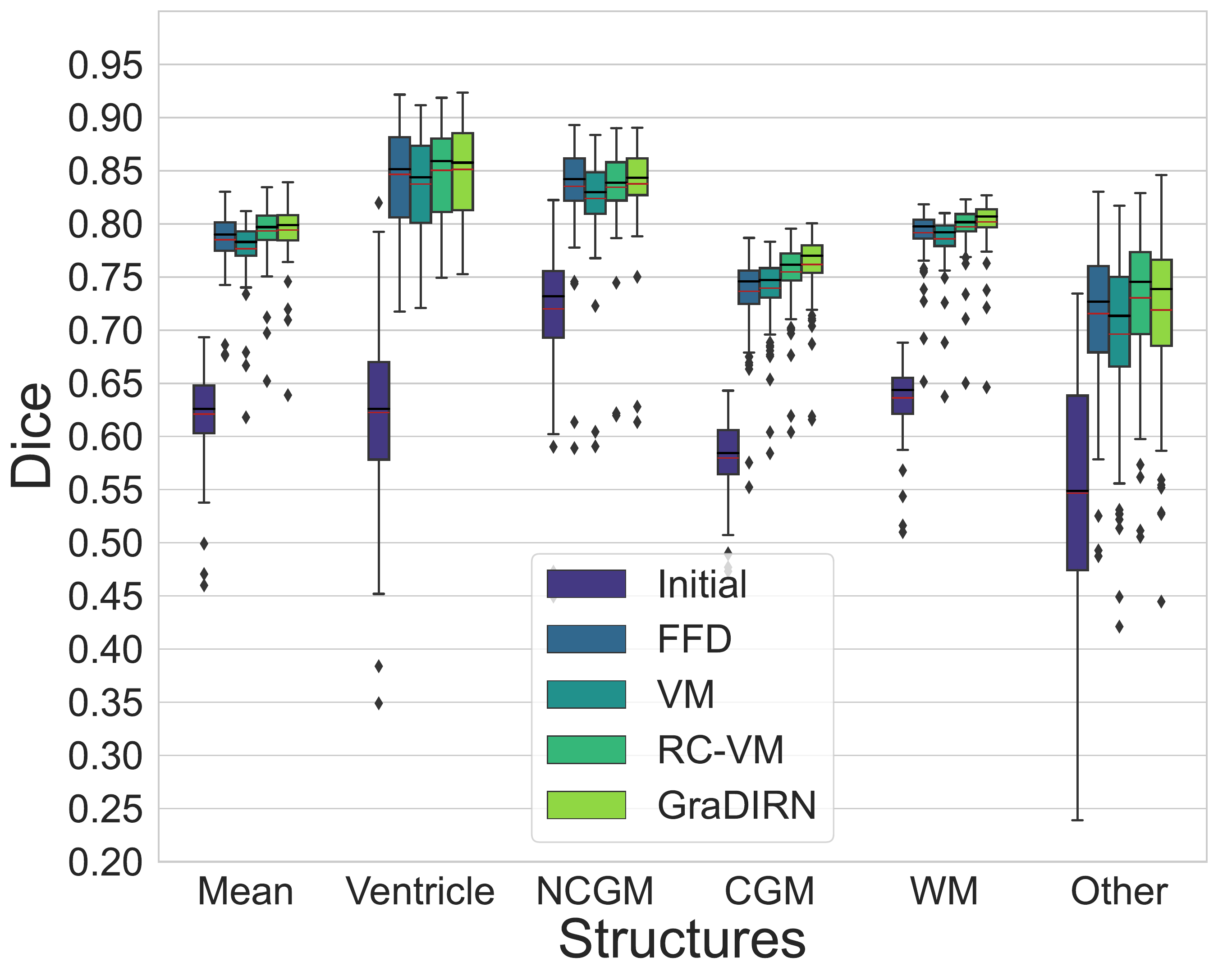}
         \caption{SSD}
     \end{subfigure}
     \begin{subfigure}[h]{0.48\textwidth}
         \centering
         \includegraphics[width=\textwidth]{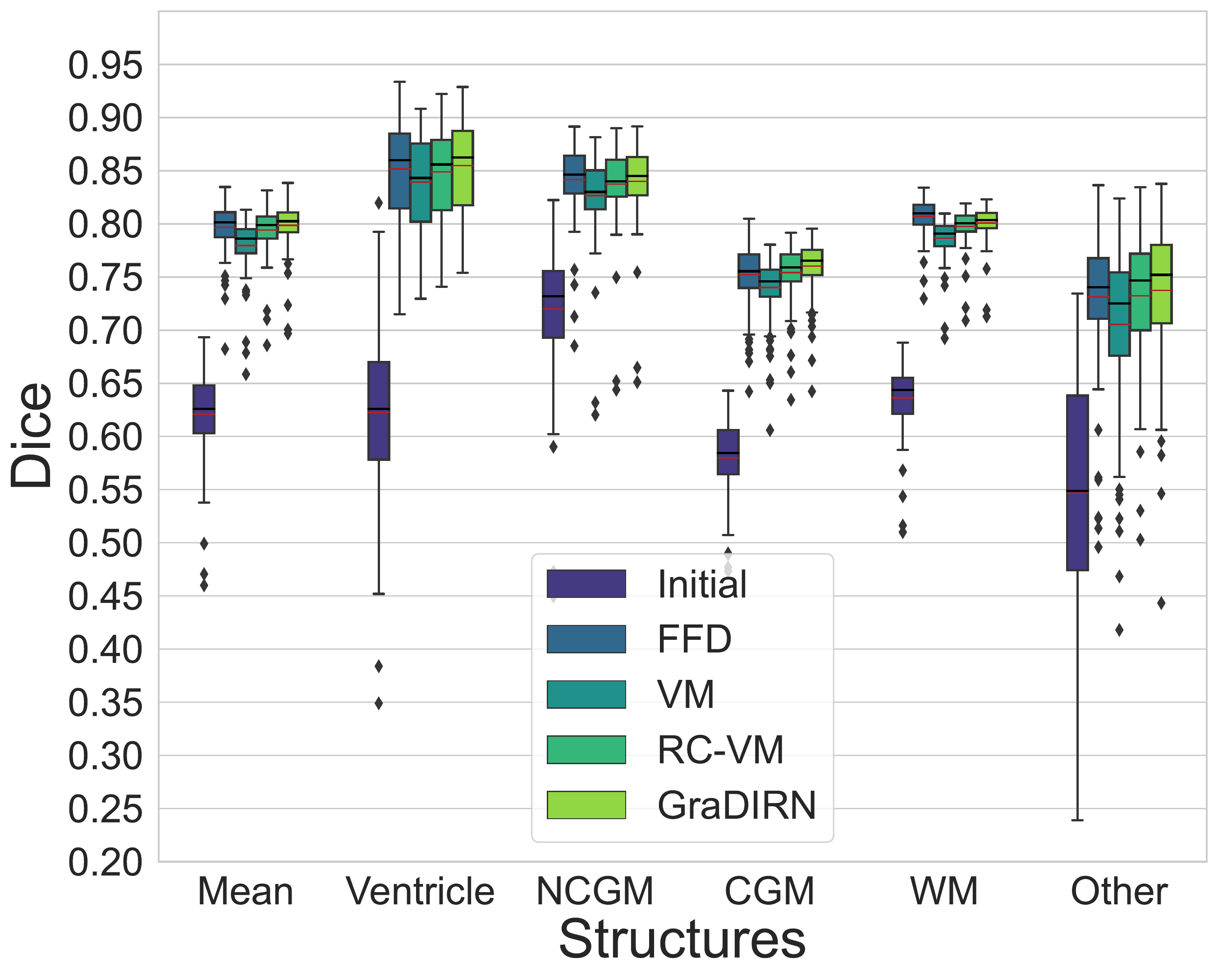}
         \caption{NCC}
     \end{subfigure}
     \caption{Boxplots of Dice scores measuring the overlap of different anatomical structures after registered by different methods in the brain registration task (CamCAN dataset), same format as \cref{fig:cardiac_boxplots}. The anaotmical structures are reported in five groups: Ventricle, Noncortical Gray Matter (NCGM), Cortical GM (CGM), White Matter (WM) and Others.}
     \label{fig:brain_boxplots}
\end{figure*}

\begin{figure}[h!]
    \centering
    \includegraphics[width=0.93\textwidth]{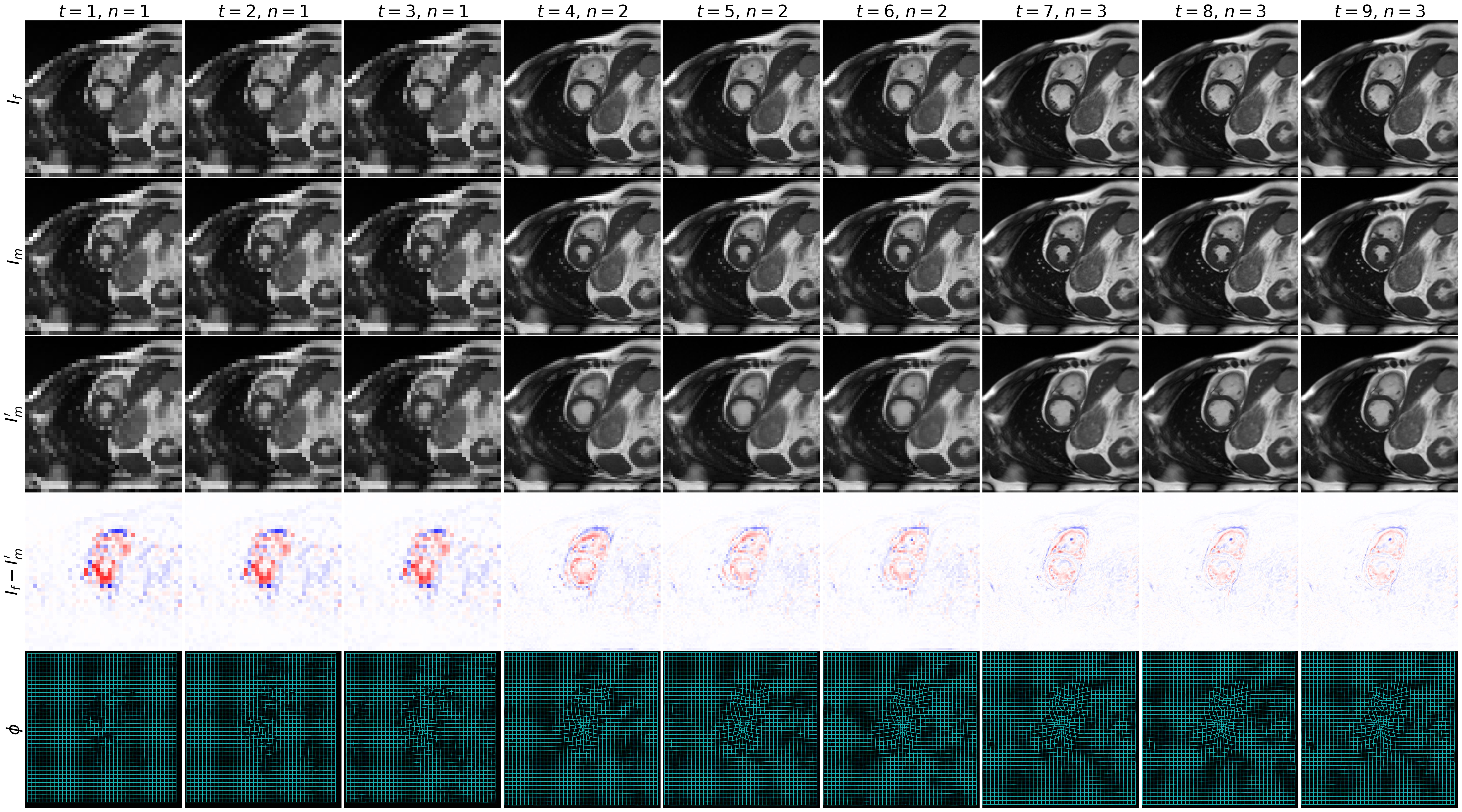}
    \caption{Illustration of the optimization steps in the forward pass of \acrshort{vn} with a cardiac MR scan from UKBB dataset. $t$ denotes step number, $n$ denotes resolution level. The transformed grids of different resolutions are resampled to the same resolution for visualization.}
    \label{fig:steps_cardiac_ncc}
\end{figure}
\begin{figure}[h!]
    \centering
    \includegraphics[width=0.93\textwidth]{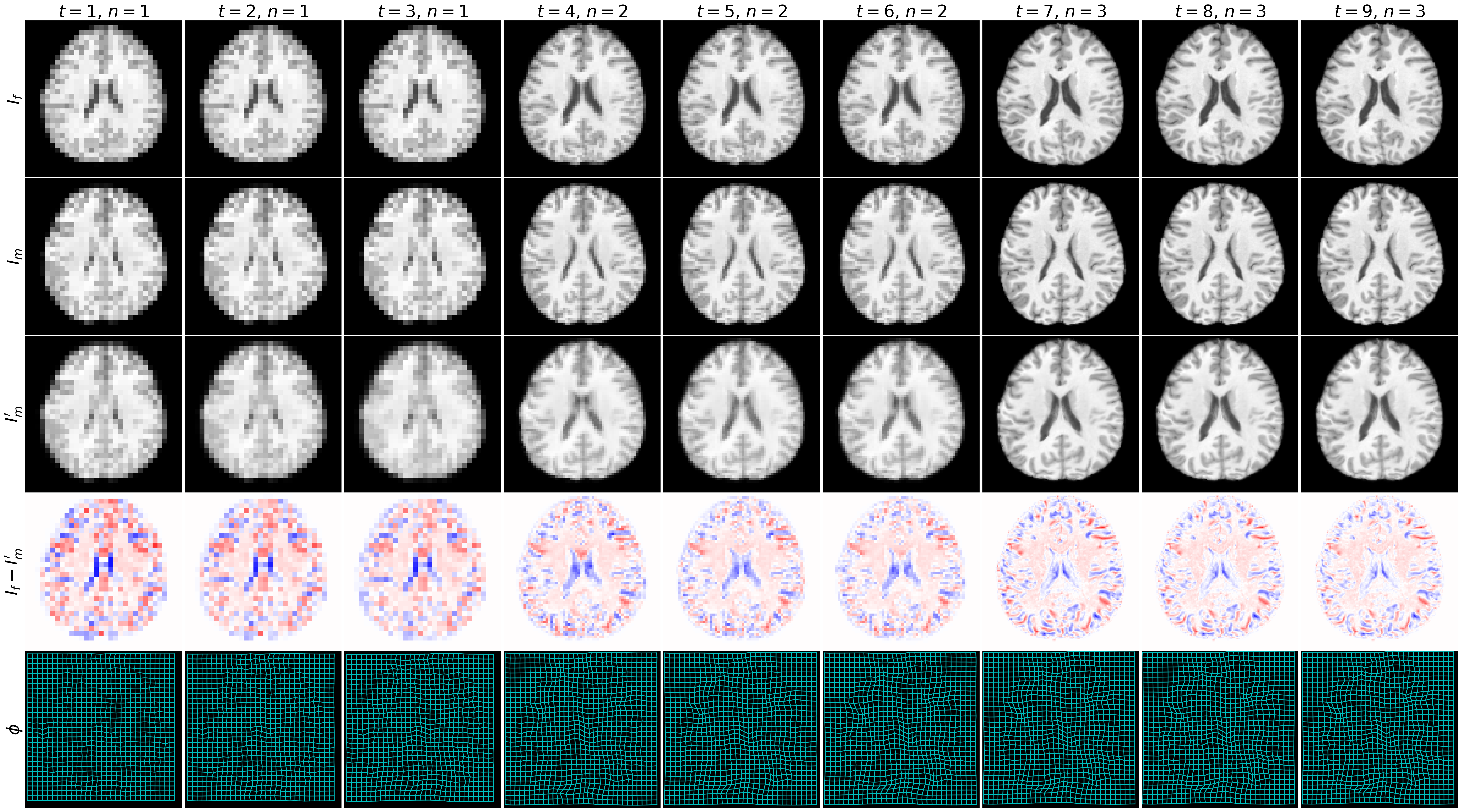}
    \caption{Illustration of the optimization steps in the forward pass of \acrshort{vn} with a brain MR scan pair from CamCAN dataset.}
    \label{fig:brain_ncc}
\end{figure}